# A Review of Temporal Aspects of Hand Gesture Analysis applied to Discourse Analysis and Natural Conversation

Renata C. B. Madeo, Priscilla K. Wagner and Sarajane M. Peres

Escola de Artes, Ciências e Humanidades, Universidade de São Paulo, Brazil
renata.si@usp.br, priscilla.wagner@usp.br, sarajane@usp.br

## ABSTRACT

*Lately, there has been an increasing interest in hand gesture analysis systems. Recent works have employed pattern recognition techniques and have focused on the development of systems with more natural user interfaces. These systems may use gestures to control interfaces or recognize sign language gestures, which can provide systems with multimodal interaction; or consist in multimodal tools to help psycholinguists to understand new aspects of discourse analysis and to automate laborious tasks. Gestures are characterized by several aspects, mainly by movements and sequence of postures. Since data referring to movements or sequences carry temporal information, this paper presents a literature review about temporal aspects of hand gesture analysis, focusing on applications related to natural conversation and psycholinguistic analysis, using Systematic Literature Review methodology. In our results, we organized works according to type of analysis, methods, highlighting the use of Machine Learning techniques, and applications.*

## 1. INTRODUCTION

With the evolution of Computer Science, including the increase of processing power and the development of new sensing devices, it has become possible – and necessary – to develop new methods capable of providing more naturalness and greater ease of use for Human-Computer Interaction (HCI). In fact, traditional methods for HCI, based on devices such as keyboard and mouse, seem to be inappropriate for some purposes, such as systems for controlling a smart home environment or systems with natural interfaces for disabled people. These applications would clearly benefit from different types of interaction, especially those that can make use of hand gesture analysis. Moreover, interaction using gestures may be combined with other modalities, such as speech and gaze, providing multimodal interaction. In recent years, this perspective has boosted the research on automated gesture analysis.

Research on hand gesture analysis is mostly focused on exploring gestures as a new way to provide systems with more natural methods for interaction. Within this scope, there are several goals to be achieved, such as developing systems which allow interaction through simple gestures, and developing systems that are capable of recognizing a predefined scope within some sign language, aiming at providing a more natural interaction for deaf people or building a communication bridge between deaf people and those who are not able to understand sign languages. Outside of this scope, automated gesture analysis may also help to build multimodal tools for specific and specialist usage, such as psycholinguistic analysis. Efforts in this direction aim at developing automated methods for feature extraction from a video stream, providing useful information for psycholinguistics researchers to analyze the video, or correlating this information to particularities of discourse, in order to automate some part of psycholinguistic analysis tasks.

In addition, to different applications for automated hand gesture analysis, there are several types of analyses that can be performed: there are studies focusing on gesture recognition, others focusing on continuous gestures segmentation (that is, the segmentation of the gesture stream into gesture units), and even studies which focus on automated extraction of some gesture feature for future application of psycholinguistic analysis. Moreover, there are other essential aspects to consider, such as strategies for acquiring data, for representing data, and for evaluating results for each type of analysis. Thus, it can be noticed that there is much research interest in gesture analysis. Within the last ten years, some reviews and surveys have been published on this subject. Such initiatives are formatted in different ways, presenting broad analysis and covering a large quantity of studies, or focusing on specific applications or methods and, consequently, covering a smaller quantity of studies. Some surveys tend to cover mostly topics which are specifically related to gesture recognition, such as [1, 2, 3]; although, more recently, the relation between gesture recognition and speech recognition has taken place in a variety of initiatives, as shown in [4, 5]. These surveys and reviews do not necessarily focus on temporal aspects and temporal tasks related to gesture analysis, such as segmenting gesture units contained in a conversation or segmenting phases within a gesture. This is an important research point justified by several definitions of gesture, which consider temporal aspects of gestures by defining them, such as a movement, motion or sequence of postures [6, 7]. Since movement is an important feature of gestures, it is adequate to analyze this feature from a temporal perspective.

Thus, we present a Systematic Literature Review (SLR) on hand gesture analysis[1], especially investigating temporal aspects. We intend to comprise applications in gesticulation and psycholinguistic analysis, besides the traditional application in predefined gestures recognition. Moreover, we aim at providing a systematic organization of several facets of gesture analysis, covering data acquisition, feature representation, methods for gesture segmentation and recognition, strategies for evaluating results, and different types of applications. Although this review do not intend to be an exhaustive study about hand gesture analysis or even a broad analysis as in [2, 4], it presents a deeper analysis of each analyzed study, classifying them according to aforementioned facets. Also, this review aims at presenting gesture analysis as a temporal task, which may be applied to several applications, including applications still ill-explored, such as gesticulation and psycholinguistic analysis. The paper is organized as follows: Section 2 gives an overview of the fundamental concepts in order to understand our analysis; Section 3 summarizes the theory of SLR methodology used in this study and describes briefly the planning and conducting phases of this review; the results analysis is presented in Section 4; and, finally, Section 5 presents some final considerations.

## 2. FUNDAMENTAL CONCEPTS

This section presents fundamental concepts and definitions of automated gesture analysis, psycholinguistic analysis and theory of gestures from Linguistics; providing readers with a minimum theoretical background to allow the better understanding of this review's content.

### 2.1. Automated Gesture Analysis

We present four concepts that will be necessary to establish the theoretical basis of the discussion presented in this review: gesture, posture, movement, and a classification of elements of gesture introduced by McNeill [8]. In this paper, these concepts are defined considering a view that meets the needs of the automation analysis area and, in some cases, they differ from view of the linguistic analysis area. These differences are briefly discussed in the text.

---

[1]From now on, the word "gesture" is used to refer to "hand gesture", unless explicitly stated otherwise.

The most significant definition for this review is the concept of "gesture" itself. It is important to notice that several authors highlight the presence of some temporal aspect in gesture definition. For instance, Corradini [6] defines a gesture as "a *series of postures* over a time span connected by **motions**". Already in [7], gesture is defined as "a **movement** that we make with a part of our body, face and hands as an expression of meaning or intention". From linguistics perspective, McNeill [8] describes gestures as **movements** of hands and arms. The same author focuses on gestures made during conversations; so, the author also highlights that these gestures are closely synchronized with the *flow of speech*. The concept "posture" is defined by Corradini [6] as "the static position of the hands and head, and the spatial relation between them". However, it is also useful to observe the hand's characteristics (fingers position or hand configuration, palm orientation and hand orientation) as descriptors of the "hand posture". Some works consider only the "hand posture" in their analysis, naming it just as "posture". A hand movement can be internal or external. Dias *et al.* [9] describe a hand movement as a bidimensional curve performed by the hand in a *period of time*, represented as the sequence of hand positions *over time*. This definition corresponds to an external movement. An internal movement is a change in hand posture or a movement made with the wrist. Since the concept of gesture is associated to movement (or motion), which can be described by temporal functions, it is appropriate to focus on temporal aspects for its analysis.

The fourth concept relates to the difference between gesticulation (that is, natural gestures which accompany speech) and gestures with predefined meaning, such as emblematic gestures (that is, a gesture with a known meaning in a specific culture, such as "thumbs-up" to indicate a job well done) or gestures from a sign language. Kim *et al.* [7] distinguish gestures into "natural" and "artificial" gestures according to their intention, stating that natural gesture is "meaningless and uncertain, and it has cultural and local diversity"; and artificial gesture "can express more detailed and various meanings using predefined motions". Wilson *et al.* [10] reinforce this idea by defining natural gesture as "the type of gesture spontaneously generated by a person telling a story, speaking in public, or holding a conversation". However, the definitions presented in [7] are not valid from linguistics perspective, since natural gestures cover all kinds of gesture that are part of a natural language and surely carry meaning. Therefore, in this perspective, gestures from sign languages, despite expressing a specific conventionalized meaning by their predefined motions, are natural gestures.

Nevertheless, in terms of automated analysis, it is relevant to distinguish gesticulation from gestures with predefined meaning. In order to be consistent with linguistic research and with automated analysis interests, we chose to use the expressions *gesticulation analysis* and *analysis of a predefined vocabulary of gestures* in this paper: *gesticulation analysis* referring to co-speech gestures analysis, i.e. the study of gestures that people make during a conversation, which meets the linguistics view and corresponds to the definition of natural gesture found in [7, 10]; and *analysis of a predefined vocabulary of gestures* covering small sets of predefined gestures aiming at HCI, or a reduced set of gestures from a sign language.

## 2.2. Psycholinguistic Analysis

Psycholinguistics is the study of the mental mechanisms behind language production and comprehension [11]. In order to study language, McNeill [8], endorsed by Kendon [12], argues that it is also necessary to consider gestures since they are part of language. In addition, these authors argue that gestures and speech are part of the same language system, arising from the same semantic source. Also, Quek *et al.* [13] reinforce that these modalities are not redundant, but co-expressive: they share a semantic source but they can express different information. The same author presents an example: while speech express that *someone has opened the door*, a two-handed unsymmetrical gesture may show *what kind of door has been opened*.

This semantic content is not shared only between speech and gesture; it may be shared among different gestures. This phenomenon is described by the concept of **catchment**, presented in [14]. A catchment occurs when two or more gesture features[2] recur in at least two gestures. This idea is based on the statement that a speaker uses gestures to represent an image in his own thinking, and if two gestures are based on the same image, it is likely that they will have some recurrent gesture features. This recurrence of images, indicated by the recurrence of gesture features, suggests a common discourse theme. According to the same author, "a catchment is a kind of thread of visual-spatial imagery that runs through a discourse to reveal the larger discourse units that emerge out of otherwise separate parts". Thus, it is possible to identify a discourse theme through detecting recurrent gesture features.

In order to understand the semantic content related to speech and gesture in a discourse, psycholinguists often perform laborious and time-consuming tasks of analyzing videos, generating very detailed transcripts. Quek *et al.* [13] described the process needed to perform a typical psycholinguistic analysis, highlighting its initial phase, called *perceptual analysis*. A perceptual analysis is made by micro-analyzing videos of subjects speaking and gesticulating. This analysis has several stages: first, speech is usually transcribed by hand and the beginning of each linguistic unit is marked with a timestamp. Second, the researcher annotates the video, marking co-occurrences of speech and gesture phases (see Section 2.3.1) and occurrences of breath pauses, speech disfluencies and other relevant comments; finally, this data is formatted into a transcript. This perceptual analysis can take about ten days to process a minute of video. Then, psycholinguistic analysis is performed using these transcripts as input.

## 2.3. Theory of Gestures

Some papers in "gesticulation analysis" intend to recognize specific types of gestures according to their temporal structure [10] or according to their semantic content [15, 16, 17]. The studies discussed in the present review are usually based on some part of gestures theory, especially on the division of gestures in **gesture phases** [12, 18] and on a taxonomy for **elements of gestures** [8]. Both are briefly commented in this section.

### 2.3.1. Gesture Phases

In this paper, we consider a framework proposed by Kendon [12] in order to describe gesture phases. This author proposes a hierarchy of gesture movements[3] applied by researchers to describe gestures. It is the case in [18], where the gesture is described as a series of different movement phases, based on such hierarchy. This framework organizes two important concepts: the gesture unit (G-unit) and the gesture phrase (G-phrase). The former is a period of time that starts when the limb moves away from the body and finishes when it comes back to rest position; and the latter occurs within a G-unit and is composed of one or more movement phases. Five phases of gestures are defined in [8, 12]: *preparation* - hands go to a position where the stroke begins; *pre-stroke hold* - a small pause at the end of the preparation, holding the position and hand posture; *stroke* - peak effort in the gesture, which expresses the meaning of the gesture; *post-stroke hold* - a small pause at the end of the stroke, holding the position and hand posture; *retraction* - return of the hand to the rest position.

---

[2]Examples: shape, location, trajectory, symmetry, and if the gesture is made by one or two hands.
[3]This framework is summarized in [8, 18].

Within a G-phrase, only the stroke is an obligatory phase; the others are optional. Also in [8], this gesture hierarchy is related to a phonological hierarchy, where a stroke corresponds to the most prominent syllable in the sentence. This relationship between gestures and phonological features is exploited in some papers, whose authors have used phonological features aiming at improving recognition of gestures [15, 16, 17]. These gesture phases are considered in some papers discussed in this review, as in [10, 18]. In [10], the authors aim at classifying gestures into biphasic and triphasic using their temporal structure; and in [18], a similar organization of gesture phases is proposed, summarized as a grammar with the following phases: an optional phase of preparation, composed by an optional liberating movement followed by a location preparation and usually blended with a hand internal preparation; an expressive phase, which may be composed by a stroke surrounded by optional dependent holds[4] or by an independent hold[5]; and an optional retraction phase. The proposed grammar helps different coders[6] to analyze signs and co-speech gestures, providing good inter-coder reliability.

### 2.3.2. Type of Gestures

McNeill [19] defines five major categories of gestures: *iconics* - associated with concrete objects or events, whose form are closely similar to an image related to the concept they represent; *metaphorics* - similar to iconic gestures, but they represent abstract ideas; *beats* - mark the rhythm of the speech[7]; *cohesives* - tie together parts of discourse that are thematically related but temporally separated, used to recall some subject after making a reference, for instance; *deitics* - also named by *pointing gestures*, they are used to indicate concrete objects and events, or even to locate references to abstract concepts in the gestural space. However, McNeill [19] argues that these categories do not represent types of gestures, but dimensions that gestures may present - a gesture may present temporal highlighting (such as beat gestures) and also be deitic (pointing at something).

## 3. Methodology: Systematic Literature Review

A systematic literature review (SLR) is a methodology to identify, evaluate and interpret the relevant research on specific topic [20]. It provides systematic methods for performing a literature review and for documenting its search strategy, allowing readers to assess its quality. The same author suggests SLRs composed of activities divided into the following phases: *planning*[8] that includes development of a review protocol, which specifies the methods to perform the review, comprising the definition of studies selection criteria and procedures; *conducting*[8] that includes selection of studies, quality assessment, data extraction and synthesis; *reporting* that includes documentation of the results in technical reports, thesis or papers.

Kitchenham [20] also describes that, in order to perform an SLR, it is necessary to perform some exploratory studies for which are not mandatory to follow methodologies. These studies are called *primary studies*, while the SLR itself is called *secondary studies*. Primary studies are necessary to know specific terms from the topic of interest, providing a foundation that helps to design the review protocol. In this paper, Section 2 provided some basic concepts needed to understand the review results, which corresponds to some results of the primary studies. The planning and conducting phases executed for this review are described in this section, and the reporting phase corresponds to the remainder of this paper.

---

[4]In [8], a *hold* means any temporary pause in the movement in the hierarchy; *rest* means the same but out of the hierarchy. A dependent hold accompanies a stroke, such as a pre-stroke hold or a post-stroke hold.
[5]An independent hold is a hold that replaces a stroke as the most significant part of the gesture – as in the gesture used for "stop" where the hand is held open in front of the body.
[6]Coder is a psycholinguistics researcher who annotates the video.
7 Beats do not carry semantic content and its semiotic value is to reveal what is important to the narrative discourse. They are also special because, unlike other types of gestures, they have only two movement phases – in/out, up/down, etc.
[8]Although the SLR phases seem to be sequential, this process is intrinsically iterative [20]: while conducting the review, the review protocol could need refinement.

## 3.1. Planning the review

The planning phase consists of identifying the need for the SLR and developing a protocol that specifies methods to conduct the review. The protocol includes specifying the review objective, research questions that should be answered by the review, strategies for studies search, data selection and data extraction. The protocol used in this SLR can be summarized by the following parameters:

- *Objective*: Identifying and analyzing temporal aspects of strategies and methods used in hand gesture analysis, especially applied to natural conversation and discourse analysis.
- *Research questions*: (1) Which types of analyses have been performed on hand gestures? (2) What strategies and methods have been applied to natural conversation and discourse analysis problems? (3) What strategies, methods and techniques have been used in temporal analyses of hand gestures?
- *Research sources*: ACM, IEEE, CiteSeerX and SpringerLink.
- *Search strings*[9]: Our objective focuses on temporal aspects of gesture analysis and the application of discourse analysis. Thus, after some pilot searches, we have chosen the following search strings. The first focuses on the application on discourse analysis; the second on temporal aspects; and the third on the specific task *continuous gesture* analysis, which always considers temporal aspects.
    1. *hand gesture* **and** (*discourse analysis* **or** *natural conversation*) **and** (*hand motion* **or** *hand movement*)
    2. *gesture* **and** (*temporal structure* **or** *gesture phases* **or** *movement phases* **or** *temporal analysis*)
    3. *continuous gesture* **and** *segment\**.
- *Idiom of the studies*: English.
- *Time restriction*: not restricted.
- *Inclusion criteria*: (I1) Studies comprising strategies, methods and techniques used in hand gestures analysis applied to natural conversation and discourse analysis; (I2) Studies comprising strategies, methods and techniques used in temporal analysis of hand gestures.
- *Exclusion criteria*: (E1) Studies which do not meet any inclusion criteria; (E2) Studies focusing on image segmentation and dataset construction; (E3) Studies focusing other kinds of gesture, not performed with the hands; (E4) Studies aiming at gesture synthesis or synthetic gestures analysis; (E5) Studies aiming at multimodal tools, with no emphasis on gesture analysis; (E6) If two papers cover the same research, the older must be excluded.
- *Studies quality criteria*: To be considered in this review, studies must be papers published in journal or conferences with peer review or thesis approved by an examining board.
- *Data selection strategy*: The search strings were submitted to the search engines. Each result was analyzed by reading its abstract considering the inclusion and exclusion criteria. If the reviewers confirmed its relevance to the SLR, the paper was fully read.
- *Data extraction strategy*: After reading the studies, the reviewers organized all relevant information about each paper: type of analysis; strategies and methods used for collecting data, feature extraction and data representation; strategies and methods used for analysis, admitting segmentation and recognition methods; results and their analysis procedures.

---

[9]The exact syntax of each string depends on the search engine.

## 3.2. Conducting the review

After defining a review protocol, it is possible to conduct the review. Table 1 summarizes included papers and the inclusion criteria met for each one.

The searches were applied at the beginning of the year 2013. Initially, searches returned 61 studies – 32 returned by ACM, 21 by IEEE, 26 by CiteSeerX and 8 by SpringerLink, with some overlapping results among the sources. After applying the inclusion, exclusion and quality criteria, 30 works were selected – 18 from ACM, 12 from IEEE, 12 from CiteSeerX and 4 from SpringerLink – resulting an acceptance rate of 49.1%.

Table 1. Included papers with Met Inclusion Criteria (MIC); exploited Type of Analysis (TA) (1 for recognition; 2 for segmentation and recognition being 2a segmentation according specific properties, 2b for segmentation using recognition, and 2c for segmentation followed by recognition; and 3 for other type of analysis); and their respective application areas.

| Study | MIC | TA | Application |
|-------|-----|-----|-------------|
| [21] | I1 | 1 | Gesticulation analysis – coreference analysis. |
| [22] | I1 | 1 | Perceptual analysis – classifying gestures as topic or speaker-specific. |
| [16] | I1 | 1 | Gesticulation analysis – recognition of beat gestures. |
| [6] | I1 | 1 | Predefined vocabulary analysis – simple gestures. |
| [23] | I1 | 1 | Predefined vocabulary analysis – simple gestures. |
| [24] | I2 | 1 | Predefined vocabulary analysis – Japanese dactylology. |
| [25] | I2 | 1 | Predefined vocabulary analysis – various datasets. |
| [26] | I1 | 2a | Perceptual analysis – sentence unit segmentation. |
| [27] | I1 | 2a | Perceptual analysis – discourse unit segmentation. |
| [28] | I1 | 2a | Gesticulation analysis – orchestra conducting gestures. |
| [10] | I1, I2 | 2a | Gesticulation analysis – classifying in biphasic and triphasic gestures. |
| [29] | I1, I2 | 2a,2c | Predefined vocabulary analysis – simple gestures |
| [30] | I2 | 2a | Gesticulation analysis – orchestra conducting gestures. |
| [31] | I2 | 2a | Predefined vocabulary analysis – aircraft marshaling gestures. |
| [7] | I2 | 2a | Predefined vocabulary analysis – gestures for controlling a smart home. |
| [32] | I2 | 2a | Predefined vocabulary analysis – simple gestures. |
| [15] | I1, I2 | 2a | Gesticulation analysis – pointing and contour gesture in Weather Narrations. |
| [33] | I2 | 2a | Predefined vocabulary analysis – simple gestures. |
| [17] | I1, I2 | 2a | Gesticulation analysis – specific gesture in Weather Narrations. |
| [34] | I1 | 2b | Perceptual analysis – oscillatory movement detection. |
| [35] | I1 | 2b | Perceptual analysis – oscillatory movement detection. |
| [36] | I1 | 2b | Perceptual analysis – hold segmentation. |
| [37] | I1 | 2b | Perceptual analysis – symmetric behavior detection. |
| [38] | I2 | 2c | Predefined vocabulary analysis – Taiwanese Sign Language. |
| [39] | I2 | 2c | Predefined vocabulary analysis – Vietnamese Sign Language. |
| [40] | I2 | 2c | Predefined vocabulary analysis – gestures for controlling slides exhibition. |
| [41] | I1 | 3 | Perceptual analysis – investigation on speech repairs. |
| [13] | I1 | 3 | Perceptual analysis – investigation on discourse structure. |
| [42] | I1 | 3 | Perceptual analysis – investigation on catchment. |
| [18] | I2 | 3 | Perceptual analysis – grammar for gesture phase segmentation. |

# 4. Results and Discussion

After the conducting phase, we executed the analysis of the studies and the review results were produced. The first analysis regards the development of gesture analysis studies through years. Our results indicate that this is not a new research area, but there has been more attention to this field in the last decade (67%). This phenomenon is probably due to advances in technology, both in areas such as ubiquitous systems, which require new methods of interaction; as well as processing power, needed in Computer Vision applications; or development of several types of sensor devices (an alternative to Computer Vision in gesture analysis systems).

Moreover, most papers (63%) have authors affiliated to institutions of United States of America. Other studies have authors of institutions from Canada, China, Japan, Republic of Korea, South Korea, Thailand, The Netherlands, United Kingdom, Spain, Italy and Vietnam. Finally, it is also possible to analyze the distribution of studies according to publication type: most papers (60%) were published in conferences, but there are also publications in journals (20%), workshops (3%), as book chapters (3%) and thesis (7%). Further analyzes are organized as follows: Section 4.1 presents applications of gesture analysis systems; Section 4.2 shows methods for data[10] acquisition; Section 4.3 presents methods for feature representation; Section 4.4 proposes a classification for dividing the studies according to type of analysis performed; methods for segmentation are described in Section 4.4.3; Section 4.4.1 lists methods for recognition; and methods for result analysis are addressed in Section 4.5.

## 4.1. Applications

In this review, we found that gesture analysis is employed for several applications. However, as we can note in Table 1, it is possible to divide most of these applications in three major groups: recognition of a predefined vocabulary of gestures (40% of included papers); gesticulation analysis (23% of included papers); and perceptual analysis (37% of included papers). This section discusses the contexts of each group of application, and highlights the temporal aspects related to them.

### 4.1.1. Analysis of a Predefined Vocabulary of Gestures

This SLR was conducted focusing on building a non-exhaustive review on different applications of hand gesture analysis, intending to cover less explored applications such as gesticulation and psycholinguistic analysis, and focusing on temporal aspects of gesture analysis. Thus, among the studies pertaining to the analysis of a predefined vocabulary of gestures, we are especially interested in how they treat the temporal aspects in the hand gestures analysis.

These papers may focus on designing a set of gestures for a specific application; or on studies aiming at sign language processing. The former studies can be used for controlling a smart home environment [7], controlling a slide presentation [40] or recognizing gestures of aircraft marshalling [31]. Some of these works also define a small set of simple gestures that could be used for HCI through gestures, such as [6, 23, 25, 32, 33]. Studies aiming at sign language processing define a reduced set of signs: 34 gestures from Japanese Dactylology [24]; 14 gestures from Taiwanese Sign Language [38]; 23 gestures from Vietnamese Alphabet [39].

However, as aforementioned, we are interested in how these studies deal with temporal aspects of hand gestures analysis. In [7, 32, 33, 38, 40], it is aimed at analyzing continuous gestures, including segmentation and recognition of a predefined set of gestures. In other studies, such as [6, 23, 25, 29, 31], it is aimed at gesture recognition and not performing temporal tasks, such as segmentation. Nonetheless, these studies consider gestures as a sequence of postures over time and incorporate this perspective into the recognition task. Corradini [6] uses Hidden Markov Model (HMM) to deal with gesture temporal structures, using Neural Networks to define HMM state transition probabilities. Spano *et al.* [23], use several descriptors (e.g. position of fingers) as a representation of a gesture at a given time, and apply this representation as a state in Non Autonomous Petri Nets in order to implement temporal gesture recognition based on states sequence analysis and model-based recognition. Yuan [25] considers gestures as a sequence of vectors, where each vector represents a frame of the gesture, and incorporates temporal machine learning algorithms, such as Recurrent Neural Networks (RNN) or Dynamic Time Warping (DTW), in order to recognize gestures. Hernández-Vela *et al.* [29], also apply an approach that represents a gesture by a sequence of frames representations, but they use Bag-of-Visual-Depth-Words organized in a histogram, which is submitted to Nearest Neighbor classifier. Choi *et al.* [31] represent each gesture as a sequence of postures obtained through K-means and uses a technique of string matching to classify sequences.

---

[10]In this case, "data" are gestures captured by cameras or sensor devices.

In a different perspective, the studies [24, 39] also work on "static gestures" or postures. Nevertheless, a static gesture (or posture) is defined, in these studies, as the main component of a gesture, similar to an independent hold, following the hierarchy described in [18]. Moreover, these studies focus on temporal aspects of gestures: Nguyen and Bui [39] segment postures (that is, the independent holds) from "transition gestures" (that is, movements such as preparation and retraction phases, which occurs between two strokes or independent holds); and Lamar *et al.* [24] use an architecture which considers a temporal normalization of gestures in order to recognize 34 gestures – 20 dynamic gestures and 14 static gestures.

### 4.1.2. Gesticulation Analysis

As defined in Section 2, gesticulation refers to gestures that people perform during conversation. Studies on gesticulation analysis usually focus on recognizing types of gestures that share some similarities regarding their structure or their semantic content, but do not necessarily share similarities regarding their shape and trajectory. Thus, a particular perspective – focusing on temporal aspects – must be considered to realize analysis in this type of gesture.

For instance, Kettebekov *et al.* [15, 17] recognize pointing and contour gestures within a Weather Narration Broadcast context. These gestures are not associated to a specific hand posture (and may even share the same posture) or to a specific trajectory, even though there may be some similarities regarding trajectory. However, pointing and contour gestures differs in their semantic content: they are performed with different intentions, conveying different messages. Both papers shows that this task may be solved using only visual cues, but a better performance could be achieved by including cues from other modalities, such as analyzing the audio signal. This performance improvement happens because both gestures and speech share the same semantic content, due to belonging to the same catchment. A similar case is seen in [16], which recognizes three types of beat gestures: transitional, which accompany discourse markers such as "but" or "so"; iconic, which are a part of an iconic gesture; and, "simple", which are the most frequent type and just mark the rhythm of speech. Such paper relies on the catchment concept considering audio signal to improve gesture recognition.

Another example is [10], in which HMM is used to model three gesture phases (rest, transition, and stroke) for recognizing gestures (distinguishing them from rest positions) and classifying them as biphasic or triphasic, considering videos of people telling a story. In this case, each class has a defined temporal structure, that is, a defined sequence of gesture phases. However, there is not a specific posture or trajectory to be associated with each class of gesture or even to each gesture phase. Such study focuses strongly on gesture temporal analysis.

Other studies on gesticulation analysis aim to analyze orchestra conducting gestures. The work [30] is similar to [10], since HMM is used in both to model gesture "phases": in [30], it is aimed at recognizing "up" and "down" movements, considering the temporal structure of orchestra conducting gestures. In [28], orchestra conducting gestures are analyzed for capturing their expressive qualities, such as "beat" and "tempo", and indicative qualities, such as "articulation". Finally, in [21], it is analyzed gesticulation to provide additional features aiming at improving coreference analysis in videos where speakers describe the behavior of a mechanical device.

### 4.1.3. Gesture Analysis for Psycholinguistic Analysis

Some researches explored in this review aim at using automated methods to extract features along videos and correlate these features with some particularities of discourse. The correlation of features extracted in an automated way may enable improving, in some aspect, part of the psycholinguistic analysis process. As described in Section 2, psycholinguistic studies are based on a perceptual analysis of video. Quek *et al.* [13] explain that, in psycholinguistic analysis, researchers analyze videos using high-quality recordings that allow performing a frame by frame analysis. As humans make this analysis frame by frame, the same author comments that it may take 10 days for performing only one minute of perceptual analysis of a video. Thus, the automation could be useful for speeding up this process.

Some studies applied to psycholinguistic analysis are developed in [13, 41, 42], which considers hand position, effort amount and holds – extracted as described in [36] – to analyze how to correlate these features to speech repairs, discourse structure and catchment, respectively. Others initiatives aim at identifying oscillatory [34, 35] or symmetric movements [37] in order to correlate these movements with discourse structure. In addition, Eisenstein [27] uses points of interests – to be described in Section 4.2 – in order to perform discourse unit segmentation and to analyze if gestures are topic-specific or speaker-specific [22]. Similarly, Chen *et al.* [26] combine a language model, a prosody model, and a gesture model aiming at detecting sentence unit boundaries. These papers can support further studies on automating tasks from psycholinguistic analysis. Thus, in these studies, temporal aspects are considered through the extraction of some features over time and through their analysis using graphics that consider these features over time. Other interesting study is [18], which proposed a grammar to help identifying gesture phases and suggested that this grammar could be used for segmenting continuous production in automatic recognition of signs and co-speech gestures. This automatic segmentation could be useful to psycholinguistics researchers since it would allow efficiently identifying gesture phases. Researchers have used the grammar proposed in [18] as basis to understand gesture phases in studies comprising gesticulation analysis [15, 17] or perceptual analysis [36].

### 4.2. Data Acquisition Method

Another important aspect of gesture analysis research pertains to methods for data acquisition. This aspect is covered in a survey [2], which states that it is possible to obtain data through sensing devices, such as instrument data gloves and body suits; or through cameras and computer vision techniques. The same study highlights pros and cons of each method: methods based on sensing devices are usually expensive and often devices are connected to a computer by cables – impairing naturalness of user interaction; while vision-based methods suffers with occlusion of parts of the user body. Regarding the papers discussed in the present review, both approaches are applied. Sensor-based approaches use data gloves [38, 39] and accelerometers [40], and vision-based approaches vary among themselves, as reported in [2].

Thus, we have divided vision-based studies in two classes, considering bidimensional (2D) and tridimensional (3D) approaches. Vision-based 2D approaches are more common and easier to execute, since they usually use only one camera to obtain an (data) image considering $x$ and $y$ axes. Most studies in this review use this approach [6, 10, 13, 15, 16, 17, 30, 31, 32, 33, 34, 37, 42]. Other studies use this approach with some particularity: a 2D approach is used in [21, 22, 24, 27], in which the users wear a colored glove, which helps to segment the object of interest in the image. A special case is [25], in which several recognition tasks with different datasets are comprised. Some datasets are obtained through 2D vision-based approaches, while other datasets are obtained using scanner devices.

Vision-based 3D approaches arise from observing that 2D features have high dependency on camera view angle, limiting systems accuracy since data acquisition may not be very accurate. Thus, the 3D approaches consist of using some method to estimate position considering $x$, $y$ and $z$ axes. In this review, we found that 3D approaches are used in [7, 26, 36, 37, 41, 42] through setting synchronized cameras in different positions, in order to estimate tridimensional positions. In [28], a 3D approach is also employed, but by estimating depth information through the use of markers - the marker size indicates depth information. A special case is [23], in which there are two different datasets: one obtained through a 3D vision-based approach based on Kinect; and other obtained using multitouch devices. Note the 2D and 3D approaches are both used in [37, 42] since two datasets are used in their analyses.

In addition to the distinction between 2D and 3D approaches, it is important to consider how data is obtained from the images. In order to obtain hands position, studies use different tracking algorithms. Several studies found in this review use Vector Coherence Mapping, which is an algorithm to extract an optical flow from a video stream, using fuzzy image processing [13, 26, 34, 35, 36, 37, 41, 42]. An algorithm that fuses tracking based on motion and skin color cues in a probabilistic framework is used in [15, 16, 17], by applying Viterbi Algorithm to determine the most probable path connecting candidates body parts over time. Eisenstein *et al.* [21] use a tracker based on Annealed Particle Filter, taking advantage of a colored glove. Lamar *et al.* [24] use tracking based on the colored glove. Corradini [6] uses a tracking method based on skin color and contour to detect hands and head positions. Swaminathan *et al.* [28] use tracking based on markers. Choi *et al.* [31] perform tracking through a Mean Shift Algorithm, which relies on the distribution of some visual features, such as color and texture. Hernández-Vela *et al.* [29] use a public dataset composed by video sequences (with RGB and depth frames), obtained by Kinect, with descriptors extraction implemented by Spatio-Temporal Interest Point detectors.

## 4.3. Feature Representation Method

Among the studies discussed in this review, there are those using only hand positions to perform an analysis [31, 34, 35, 37, 40], and those building other features based on hand position. For instance, Bryll *et al.* [36] use a sliding window approach to compute an amount of effort concerning hand position in order to determine if the segment is part of a hold or not. Chen *et al.* [26, 41] consider hand positions and values proposed in [36]: amount of effort and presence of hold. Quek *et al.* [13] and Quek [42] consider all previous characteristics and the presence of symmetry, extracted as described in [37]. Werapan and Chotikakamthorn [38] and Nguyen and Bui [39] use hand position to derive velocity features. Kettebekov *et al.* [15, 17] and Kettebekov [16] use hand position to derive velocity and acceleration features. Corradini [6] considers the angle between hands and head and derives other features from velocity. Swaminathan *et al.* [28] consider velocity and direction of the movements. It is interesting to note that all studies focusing on supporting psycholinguistic analysis tasks use only features based on hand position information, such as coordinates, amount of effort, and presence of holds or symmetry. A special case is [23], in which is used, depending on the application, position of fingers or of skeleton joints, joints orientation and hands states to represent the gesture in a given time (a state). Thus, a gesture model is represented through a sequence of states.

There are other approaches to represent a gesture without obtaining the hands or head positions, but using the whole images in the frames that compose the gesture recording. Wilson *et al.* [10] compute the eigenvector decomposition of each frame, extracting 10 coefficients to represent each frame; Eisenstein *et al.* [22] and Eisenstein [27] extract spatiotemporal interest points with an Activity Recognition Toolbox; Eisenstein *et al.* [21] extract some features regarding position and speed, some features based on clustering segments into rest and movement segments, and some features comparing two gestures, such as measure of agreement between two gestures using DTW; Wong and Cipolla [32] compute a Motion Gradient Orientation from the whole image for each frame; and Yuan [25] uses Hu Moments to extract coefficients from the whole image. Other studies focus on the user silhouette, for example, Wilson and Bobick [30] extract the spatial distribution of silhouette pixels using a tracking algorithm based on Expectation-Maximization (EM); Kim *et al.* [7] represent the silhouette contour by 80 points and uses Principal Component Analysis to reduce dimensionality; and Li and Greenspan [33] represent the silhouette contour with a border signature. Finally, Hernández-Vela *et al.* [29] use Bag-of-Visual-and-Depth-Words organized on histograms with the frequencies of such visual words, which are chosen among several possible image descriptors, by applying k-Means algorithm.

## 4.4. Type of Analysis

Observing our results, it was possible to identify three major types of gesture analysis: gesture recognition (1), considering previously segmented gestures (23.3% of included papers); gesture segmentation and recognition (2), which considers a video with continuous gesture, segments the gesture, and recognizes it, or perform some kind of segmentation using recognition methods (63.3% of included papers); and other type of analysis (3), including mainly perceptual analysis (13.3% of included papers).

Most papers discuss some kind of segmentation and recognition. Among these papers, it is possible to identify three subdivisions: studies that perform segmentation based on the results of a recognition model (2a); studies that perform segmentation and recognition simultaneously, usually according to some specific property, such as symmetry or oscillatory gestures, to segment sections that contain that property (2b); and studies that first perform segmentation for later recognition (2c). Table 1 shows the classification for each included study. In order to support the analysis of the papers explored in this review, the remainder of this section clustered them as follows: static gesture recognition task; gesture recognition task solved by Machine Learning techniques; gesture segmentation strategies based on recognition models, based on filters or thresholds and based on turning points detection; and finally, an explanation about the relation between segmentation and recognition tasks.

### 4.4.1. Recognition in Previously Segmented Gesture

Studies that perform only recognition, with no segmentation task, usually aim at recognizing gestures considering that these gestures have already been segmented, that is, these studies do not intend to perform continuous gesture recognition. Therefore, these studies use previously segmented gestures both to compose the training and the testing sets used in the recognition model construction. This approach is used in [6, 23, 24, 25] to recognize sets of simple gestures acquired in an individual way. In [16], which is applied to gesticulation analysis, this strategy where gestures are segmented manually using special software is used. In other study [21], the video segments representing a gesture are defined manually using speech to establish which portions of the video will be analyzed, since it uses only gestural cues to help analyze coreferences in speech. Other approach consists in performing no segmentation. In [22], which is applied to psycholinguistic analysis, short videos containing dialogues are analyzed to extract features and define if these features are topic-dependent or speaker-dependent. Thus, there is no need of segmenting each gesture, since features are analyzed in the videos as a whole.

### 4.4.2. Recognition Methods

Regarding recognition methods themselves, most studies use a Hidden Markov Model (HMM) – in 29% of all studies that perform some recognition task. Typically, HMM methods are used when there is a predefined set of gestures to be recognized, as in [40]. However, this review has found that they have also been used to recognize more abstract classes of gestures, as in [15, 16, 17]. Chen *et al.* [26] use different Machine Learning methods to build three different recognition models: *language model* using a statistical language model; *gesture model* using Bayesian Decision Tree or CART Decision Tree; *prosody model* using Bayesian Decision Tree. Then, the authors apply a HMM in order to integrate the answers from these modules and obtain a final decision on the detection of sentence boundaries.

Another interesting use of HMM is made in [30]. It uses HMM to determine only gesture temporal structure, but the training for associating the user image to each HMM phase is performed at the beginning of the interaction with the system rather than in an offline way. A similar approach is used in [10], which models gesture temporal structure with a Finite State Machine (FSM) and uses this information to classify gestures in biphasic or triphasic. Moreover, HMM is combined to RNN in a hybrid architecture in [6], which uses Neural Networks to determine the transition probabilities from one HMM state to another.

Other recognition methods are also studied. Lamar *et al.* [24] use a neural network strategy, called T-CombNet. A T-CombNet is composed by a Learning Vector Quantization neural network as a first layer in order to cluster similar data and direct this similar data to branches of RNN, where classification takes place. Corradini [6] combines HMM and Neural Networks, using Neural Networks to learn probabilities for states transitions. Yuan [25] also uses Neural Networks and several other approaches, such as Support Vector Machines (SVM), Dynamic Time Alignment (DTA) and DTW, aiming at recognizing several sets of gestures. Li and Greenspan [33] use Continuous Dynamic Programming (CDP), and Choi *et al.* [31] employ K-means associated with Longest Common Subsequence technique. Spano *et al.* [23] propose a compositional, declarative meta-model based on Petri Nets for recognizing gestures. Swaminathan *et al.* [28] use a Hypotrochoidal Model (HP), a Bayesian framework, and a Switching State-Space model (SSM), aiming at modeling beat, tempo, and recognizing articulation. Hernández-Vale *et al.* [29] apply k-Means algorithm in order to organize the gesture representation, DTW to segment the gesture, and Nearest Neighbor classifier in order to carry out the gesture recognition. There are also studies with probabilistic approaches, using Bayesian Analysis, as in [22, 27]; or using Relevance Vector Machine, as in [32]. Table 2 summarizes the recognition methods used in each paper.

### 4.4.3. Strategies for Segmentation

Regarding studies that comprise gesture segmentation, it is possible to identify three strategies: performing segmentation according to the probability of existence of a gesture; applying some kind of filter or threshold to all frames; and detecting a turning point in the signal, which usually indicates the beginning or end of a gesture. Table 2 lists the segmentation strategy addressed by each study, according to the classification defined, and discussed, in this section.

**4.4.3.1. Segmentation based on recognition model**

Segmentation based on gesture recognition is typically used for continuous gesture analysis. In such cases, it is necessary to segment each gesture belonging to a sequence and then analyze the segmented gestures. Thus, some studies estimate the probability of existence of a gesture, by implementing a probabilistic recognition model, and use these models to segment gestures.

Wong and Cipolla [32] establish classes of gestures and generate hypothesis about the beginning and the end of a gesture, suggesting a video segment. Then, the authors apply Relevance Vector Machines (RVM) to estimate the probability of existence of all classes in that segment of video. When the probability of existence of any class exceeds a certain threshold, the segment is defined as valid. Already in [33], CDP is used to detect segments where some gesture probably exists. Usually, classification using CDP relies on finding the minimum accumulated distance value to a pattern, in this case "a gesture", since CDP creates a reference model for each class of gesture and measures the distance between reference model and each input. Then, the video sequence is scanned for gestures, that is, sequences that have a small distance from some reference model. In [29], DTW is used to detect gestures of reference that provide splits of multiple gestures to be recognized in a next step. Also, Eisenstein [27] uses a similar approach for discourse topic segmentation. In this study, a Bayesian classifier model is developed to evaluate the likelihood of a specific segment being a cohesive topic in the discourse. Then, a set of segments is evaluated and the best segmentation is chosen according to the probabilities found by the Bayesian framework. Similarly, Chen *et al.* [26] use classifiers – a statistical language model and decision trees – aiming at detecting sentence unit boundaries. This study considers gestural, lexical and prosodic information in order to classify each segment as a sentence unit boundary or not. Thus, the video is segmented according to the boundaries found by the classifiers. Another similar method used in [15, 17] is called Token Passing. This method is based on Viterbi decoding and calculates iteratively the likelihood of possible sequential gesture interpretations for each segment. Thus, again, segments are defined according the probability of occurrence of each gesture.

Table 2. Segmentation and recognition methods used in each study

| Study | Segmentation Method | Recognition Method |
|-------|---------------------|--------------------|
| [21] | — | Conditional Modality Fusion |
| [22] | — | Bayesian Model |
| [16] | — | HMM trained with Baum-Welch algorithm |
| [6] | — | Hybrid architecture with RNN and HMM |
| [24] | — | T-CombNet (Neural Network model) |
| [25] | — | SVM, DTA and DTW; CDP |
| [26] | Recognition-based segmentation | Bayesian Decision Tree |
| [27] | Recognition-based segmentation | Bayesian Analysis |
| [29] | Recognition-based segmentation | DTW, Nearest Neighbor |
| [10] | Recognition-based segmentation | FSM |
| [30] | Recognition-based segmentation | HMM trained with Baum-Welch algorithm |
| [31] | Recognition-based segmentation | K-means with Longest Common Subsequence |
| [32] | Recognition-based segmentation | Relevance Vector Machine |
| [15] | Recognition-based segmentation | HMM |
| [33] | Recognition-based segmentation | CDP |
| [17] | Recognition-based segmentation | HMM |
| [28] | Recognition-based segmentation | HP, Bayesian framework, and SSM |
| [39] | Threshold-based segmentation | Based on hand velocity |
| [36] | Threshold-based segmentation | Heuristic based on effort amount calculating |
| [37] | Threshold-based segmentation | Heuristic – correlation between hand motion |

| [35] | Filter-based segmentation | CWT |
|---|---|---|
| [34] | Filter-based segmentation | WFT and CWT |
| [38] | Segmentation based on turning point detection | Based on hand acceleration |
| [40] | Segmentation based on turning point detection | HMM trained with EM |
| [7] | Recognition-based segmentation combined with turning point detection | Accumulative HMM with voting scheme |

The studies in [10, 30] use a FSM and a HMM, respectively, to process gestures. The former aims at classifying biphasic and triphasic gestures. For achieving this purpose, it builds a FSM with three states corresponding to gesture phases: rest ($R$), transition ($T$ – corresponding to preparation or retraction) and stroke ($S$). The same paper considers that triphasic gestures have a structure in the form $< R - T - S - T - R >$, while biphasic gestures can be represented as $< R - T - R >$. Thus, this study performs an implicit segmentation into gesture phases within FSM functioning. Also, there is an explicit segmentation of the video based on the FSM response in order to divide the video into periods with biphasic gestures, with triphasic gestures, and with no gesture. The latter study consists in an application where the user performs orchestra conducting gestures and his movements are the input in order to control the tempo of a music played by the application. In this case, a HMM with three states – rest, up, and down – is used. The segmentation is also based in the transition of the HMM states: depending on the time between state changes, the application answers by changing the tempo of the song. Thus, this papers segments a gesture considering HMM state changes or even HMM classification responses.

Swaminathan et al. [28] use its recognition model – based on HP, Bayesian framework, and SSM – for segmenting strokes of conducting gesture. Choi et al. [31] perform a clustering, considering each frame as belonging to a cluster of postures, and perform recognition using Longest Common Subsequence (LCS), i.e., gestures are represented by sequences of postures, and the longest valid sequence is considered as the actual sequence. Since this method considers the longest valid sequence of gesture, the authors indicate that it may be used for segmentation.

### 4.4.3.2. Segmentation based on filters or thresholds

Another approach consists in applying some kind of filter or threshold to all frames. This strategy is used in [39]. This study aims at segmenting static gestures from Vietnamese Sign Language by analyzing hand velocity. The presence of a static gesture is verified when hand velocity is below a certain threshold, i.e., a set of frames, in which velocity remains low, contains a static gesture segment. Else, such set of frames contains a segment that is considered as a transition movement. Some studies aim at recognizing some property in the gestures and segmenting sections with the desired property in the video. Among the papers found in this review, this kind of study usually employs some kind of filter or threshold in order to simultaneously segment and recognize the desired property, such as symmetry, oscillatory movements and periods of hold. Following this line, Xiong et al. [37] calculate a correlation between right hand and left hand signals, and when such correlation exceeds a threshold, segments with symmetry are detected. Similarly, Xiong and Quek [34] apply the filters Windowed Fourier Transform (WFT) and Continuous Wavelet Transform (CWT) to the signals, represented by hand positions along the video, and extract the signal peaks to detect oscillatory movements. A similar strategy is used in [35], but only with CWT. Also, Bryll et al. [36] aim at detecting holds by employing an adaptive threshold using "dominant hand rule", which declares that the amount of movement necessary to define that a hand is in hold depends on the amount of movement performed by the other hand. Then, a temporal filtering eliminates hold and no-hold periods considered too short.

### 4.4.3.3. Segmentation based on detecting a turning point

The third strategy consists of detecting a turning point in the gesture sequence, which usually represents the beginning or end of a gesture. This approach is used in [38, 40]. The former searches for segments where movement velocity reaches a minimum, inferring the moment when the movement should be starting or ending. The latter performs segmentation based on the hypothesis that acceleration changes more abruptly at the beginning or end of a gesture.

### 4.4.3.4. Combining strategies

Moreover, the strategies aforementioned could be used together. For instance, Kim *et al.* [7] calculate a Competitive Differential Observation Probability (CDOP – the difference between the maximum probability of existence of some gesture and the probability of having no gestures for each frame), which corresponds to the first strategy; then, CDOP is used to determine turning points: when CDOP turns from negative values to positives values (or vice-versa), it is likely to be the beginning or the end of a gesture.

### 4.4.4. Relation between segmentation and recognition tasks

Some studies perform segmentation by using recognition techniques to resolve the probability of containing a gesture to each segment (2a). These studies mostly rely on a dataset of previously segmented gestures used for training, as in [16, 32, 33]. Other studies perform segmentation and recognition simultaneously, usually by employing some heuristic method (2b). It is the case of studies that perform segmentation according to the detection of some property, as described in Section 4.4.3 for the papers [34, 35, 36, 37]. In this case, the recognition is performed by preprocessing these signals and applying thresholds. A specific case happens in [29], where it is implemented an approach with two parts: first, DTW is used as a recognition model that solves a segmentation task; then the previously found out segments are used as input for a recognition model capable to decide what gesture is been performed.

The strategies described above are directly related to segmentation strategies. However, the third strategy does not strongly correlate segmentation and recognition strategies, since it aims at performing segmentation by recognizing turning points based on information about hand velocity and acceleration to select the most probable points where gestures may begin or end, and later performing recognition (2c). Werapan and Chotikakamthorn [38] consider velocity for selecting probable turning points. Then, they analyze the signal using Fourier analysis and searching spectrum peaks for detecting periodic movements – since in Thai Sign Language signs are usually composed by periodic movements, then non-period movements are less likely to be signs. Thus, in this paper, recognition corresponds to the detection of periodic movements. Similarly, both in [39, 40], segmentation is carried out by selecting probable turning points, and then recognition is performed using Machine Learning techniques (Section 4.4.2).

## 4.5. Methods for Result Analysis

Most studies performing recognition tasks focus on recognition rate as the metric for result analysis [7, 27, 31, 38, 40]. However, other statistic measures and analysis methods are also used for studies focusing on recognition tasks: Kettebekov [16] includes a confusion matrix; Lamar *et al.* [24] discuss, in addition to recognition rates, the number of neurons and time consumed by its approach; Nguyen and Bui [39] present statistics of precision and recall; Bryll *et al.* [36] also show statistics of precision, accuracy, recall and specificity; and Eisenstein *et al.* [21] use Area Under ROC (Receiver Operating Characteristic) Curve.

On the other hand, studies focusing on continuous gesture recognition, which include segmentation, may use a recognition metric originated from speech analysis research that considers three types of error: insertion, deletion and substitution. An insertion error occurs when an extra gesture, which does not exist, is recognized. A deletion error occurs when an existing gesture is not recognized. A substitution error occurs when an existing gesture is recognized as a different gesture, that is, the gesture is classified as being from a wrong class. Such metric was used and briefly described in [15, 17, 29, 32, 33]. Chen *et al.* [26] also use this metric, but in an adapted way: as this study aims at recognizing sentence boundaries, it is only possible to have insertion errors, when a sentence boundary is detected in a place that has no boundary; and deletion errors, when a true sentence boundary is not detected.

Another strategy used for evaluating results of gesture segmentation and recognition includes the comparison with human labeled videos. Wilson *et al.* [10] present their results through a graphic comparing the labels assigned by the automated method versus human labeling. Nevertheless, it is important to note that human labeling may be inaccurate: Kita *et al.* [18] show that agreement between human specialists in the task of defining gesture phases may be improved by using a well-defined grammar, but it still does not allow very high levels of agreement (72% for gesticulation and 58% for sign language). Thus, evaluating an automated method with a human coder should include a comparison between different human coders, allowing the assessment of differences between inter-coder variations and variations between the coders and the automated method.

Some studies in this review also rely only in graphics for showing the results. Xiong and Quek [34] analyze graphics showing the results of applying CWT and WFT to hand location signals. It is through these graphical analyses that the author argues that CWT is more appropriate for detecting oscillatory movements. Also, Swaminathan *et al.* [28] use graphics to presents results on inferring the type of articulation – *legato* or *staccato* – in a given segment of music. In [23], the analysis focused on the quality of the logical solution that implements gesture recognition. So, the authors used simplified experiments in terms of complexity of the gestures, and in such contexts, the gesture recognition problem was completely well-solved.

Moreover, there are papers [13, 35, 37, 41, 42] which focus on psycholinguistic analysis, using automated extracted features as part of the analysis and trying to correlate these features with discourse structure, for instance. Xiong and Quek [34] also perform a psycholinguistic analysis – after detecting oscillatory movements using CWT – in order to correlate the presence of oscillatory movements to changes in discourse structure. The results in these papers are produced through psycholinguistic analysis itself, which consists in qualitative techniques of results evaluation used in disciplines such as psychology, psycholinguistics, and anthropology.

## 3. FINAL CONSIDERATIONS

This paper presented an SLR of research on hand gesture analysis, especially by analyzing its temporal aspects. As SLR requires a formal methodology, this paper summarized its parameters through describing the protocol and the conducting process. Then, the results of this systematic research were analyzed according to some topics, such as type of applications and strategies for data acquisition, feature representation, type of gesture analysis, and results evaluation.

Regarding types of applications, this review has proposed a classification in three major groups of applications: analysis of a predefined set of gestures, gesticulation analysis, and psycholinguistic analysis. It is important to notice that the latter two types of applications have not been focus of previous surveys on gesture analysis. Thus, this paper highlighted these types of applications, even indicating some topics for future research in automated analysis aiming at developing tools for supporting psycholinguistic analysis. We also presented strategies for data acquisition, including sensor and vision-based approaches, and feature representation, most of them based on features derived from hand position. Type of analysis was divided in three classes: studies comprising only gesture recognition, studies comprising gesture segmentation and recognition, and other types of studies, including psycholinguistic analysis. Concerning gesture segmentation, we have shown some strategies and have classified them in three groups: segmentation approaches based on recognition, approaches based on filters and thresholds, and approaches based on detecting a turning point. Regarding gesture recognition, we have presented the application of some Machine Learning methods and some other approaches for gesture recognition. Finally, we have enumerated some strategies for evaluating results.

This review has the following differentials with respect to other reviews and surveys on gesture analysis: analyzing temporal aspects and tasks of gesture analysis; highlighting different types of applications, such as gesticulation analysis and tools for supporting psycholinguistic analysis; and providing a systematization on several topics, such as types of applications and strategies for data acquisition, feature representation, type of analysis, and results evaluation. On the other hand, this review has some limitations pertaining to reliability about subjective factors related to the conducting process of the SLR and about the extent to which literature has been covered. Based on primary studies, we established *search strings* and *inclusion and exclusion criteria*. The application of these criteria is a subjective task and different researchers could reach similar results, but not necessarily identical results. However, despite such limitations, we considered that this review is suitable to answer three research questions mentioned in Section 3: what type of analyses have been performed on hand gestures; what strategies, methods and techniques have been applied to natural conversation and discourse analysis problems; and what strategies, methods and techniques have been used in temporal analyses of hand gestures. Although we have not covered all research already conducted on these topics, the coverage of our review protocol allowed us to propose interesting classifications for several aspects related to gesture analysis research.

Both the discussion presented in this review and the applied review protocol could be refined in future works. Some possibilities to further contributions are: (a) extending the search for studies considering other research sources, including digital libraries correlated to areas such as Linguistics and Psychology, where studies that can provide formal bases for automated gesture analysis are developed; (b) accomplishing an SLR conducting process in interdisciplinary way, involving researchers from Computer Science and Linguistics, in order to minimize the effects of subjective analysis taking into account the interests of both research communities.

## ACKNOWLEDGEMENTS

The authors thank São Paulo Research Foundation (FAPESP) - Process Number 2011/04608-8; and Tutorial Education Program of Ministry of Education, Brazil.

**Authors**

Renata Cristina Barros Madeo is Master of Science from the Graduate Program in Information Systems, at the School of Arts, Sciences and Humanities, University of São Paulo (USP), Brazil. Her main research interests are computational intelligence and pattern recognition.

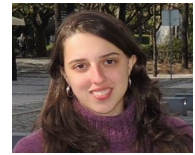

Priscilla Koch Wagner is an under graduating student in the Under Graduate Program in Information Systems, in the School of Arts, Sciences and Humanities, University of São Paulo (USP), Brazil. Her main research interests are computational intelligence and pattern recognition.

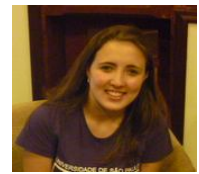

Sarajane Marques Peres is an assistant professor in the School of Arts, Sciences and Humanities, University of São Paulo (USP), Brazil. She earned a PhD in electrical engineering (2006) at the University of Campinas (Unicamp) and a ME (1999) at the Federal University of Santa Catarina (UFSC), Brazil. She has worked as researcher-professor at State University of Maringá and Western Paraná State University. Her main research interests are computational intelligence and pattern recognition.

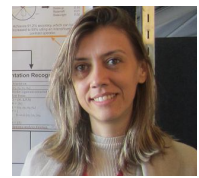